\documentclass{jfm}

\usepackage{mathpazo}
\usepackage{amsmath}
\usepackage{ucs}
\usepackage{array}
\usepackage{ifthen}
\usepackage[utf8x]{inputenc}

\usepackage{pdfpages}
\usepackage[normal]{caption}
\usepackage{verbatim}
\usepackage{multicol}
\usepackage{multirow}
\usepackage{wrapfig}
\usepackage{float}
\usepackage[T1]{fontenc}
\usepackage{graphicx}
\usepackage{subfigure}
\usepackage{verbatim} 
\usepackage{upgreek}
\usepackage{wasysym}
\usepackage{numprint}
\usepackage{mathrsfs}
\usepackage{tikz}
\usepackage{textcomp}
\usepackage{appendix}
\usepackage{pifont}
\usepackage{threeparttable}

\usepackage{amssymb}
\usepackage{tabularx}
\usepackage{hyperref}
\usepackage{hypcap}
\usepackage{ulem}
\usepackage{color}

\usepackage{natbib}

\usepackage{pifont}
 
\ifCUPmtlplainloaded \else
  \checkfont{eurm10}
  \iffontfound
    \IfFileExists{upmath.sty}
      {\typeout{^^JFound AMS Euler Roman fonts on the system,
                   using the 'upmath' package.^^J}%
       \usepackage{upmath}}
      {\typeout{^^JFound AMS Euler Roman fonts on the system, but you
                   dont seem to have the}%
       \typeout{'upmath' package installed. JFM.cls can take advantage
                 of these fonts,^^Jif you use 'upmath' package.^^J}%
      }
  \else
  \fi
\fi

\ifCUPmtlplainloaded \else
  \checkfont{msam10}
  \iffontfound
    \IfFileExists{amssymb.sty}
      {\typeout{^^JFound AMS Symbol fonts on the system, using the
                'amssymb' package.^^J}%
       \usepackage{amssymb}%

      }{}
  \fi
\fi

\ifCUPmtlplainloaded \else
  \IfFileExists{amsbsy.sty}
    {\typeout{^^JFound the 'amsbsy' package on the system, using it.^^J}%
     \usepackage{amsbsy}}
    {\providecommand\boldsymbol[1]{\mbox{\boldmath $##1$}}}
\fi

\title{Finite-size effects in parametric subharmonic instability}
\author[B. Bourget, H. Scolan, T. Dauxois, M. Le Bars, P. Odier, S. Joubaud]{Baptiste Bourget$^1$, Hélène Scolan$^1$, Thierry Dauxois$^1$, Michael Le Bars$^{2}$, Philippe Odier$^1$, Sylvain Joubaud$^1$
}
\affiliation{
1. Laboratoire de Physique de l'\'Ecole Normale Supérieure de Lyon, Université de Lyon, CNRS, 46 Allée d'Italie, F-69364 Lyon cedex 07, France.\\
2. CNRS, Aix-Marseille Université, Ecole Centrale Marseille, IRPHE UMR 7342, 49 rue F. Joliot-Curie, 13013, Marseille, France.}

\date{\today}

\begin{document}
\maketitle
\begin{abstract}
The  parametric subharmonic instability in stratified fluids depends on the  frequency and the amplitude of the primary plane 
wave. In this paper, we present experimental and numerical results emphasizing that the finite  width of the beam also plays 
an important role on this triadic instability. A new theoretical approach based on a simple energy balance is developed and 
compared  to numerical and experimental results. Because of the finite width of the primary wave beam, the secondary pair 
of waves can leave the interaction zone which affects  the transfer of energy. Experimental and numerical results 
are in good agreement with the prediction of this theory, which brings new insights on energy transfers in the ocean where internal waves with finite-width beams are dominant.
\end{abstract}


\smallskip

\section {Introduction}
 Nonlinear resonant interaction of internal waves is one of the key processes leading
 to small-scale mixing in the ocean. In particular the parametric subharmonic instability (PSI) allows the transport of energy from large to smaller scales by giving birth to two secondary subharmonic waves (with wave vector modulus $\kappa_1$ and $\kappa_2$) from an initial primary wave (with wave vector modulus $\kappa_0$)~\citep{BenielliSommeria1998,KoudellaStaquet2006,Bourget2013,Gayen2013,Clark2010}.
Thanks to a new experimental and analysis tool, \cite{Bourget2013} have recently shown how the PSI theory developed for plane waves is in good agreement with the experimental observations of the instability of a quasi-monochromatic wave beam.
As stressed by~\cite{Sutherland2013}, one challenge is now to determine the range of validity of the theory and in particular  the role of the width of the wave beam on the occurrence of the instability. In other words, is the instability affected by the width of the beam? This question is particularly important in the oceanic context where internal waves are known to develop preferentially in the form of finite size beams, for instance waves emitted by the interaction of tide with topography~\citep{Lien2001,Dewan1998,Gostiaux2007b}. Moreover, energy transfer between scales, as well as wave turbulence in oceans, are open questions and a study of the role of the beam width on secondary waves selection in PSI might bring new understanding to these issues.

 This is the aim of the present work, combining experimental, numerical and theoretical approaches. Starting with the observation that finite-width beams can inhibit the instability, we continue by studying the effect of the beam width on the selection rules.

\section{Experimental and numerical approach}
\subsection{Experimental set-up.}
\begin{figure}
\begin{center}
\includegraphics{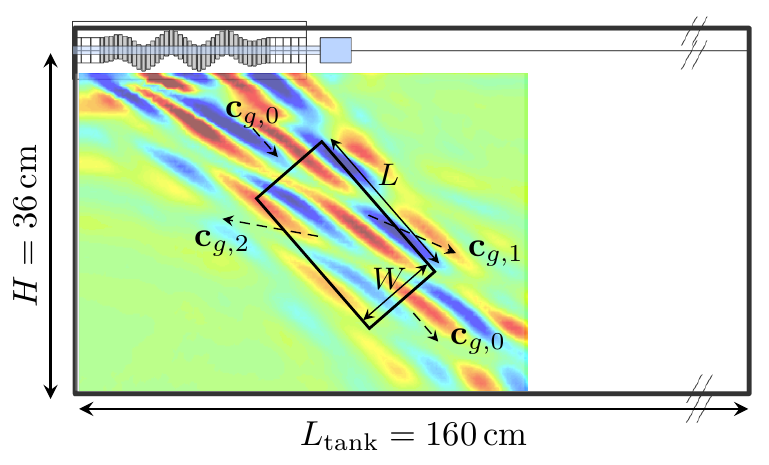}
\caption{Sketch of the set-up. Experimentally, the wave generator is lying horizontally at the top of the wave tank.
The vertical density gradient field of a typical wave beam generated at the top of the domain and undergoing PSI is presented as background of the figure. The dashed arrows indicate the group velocity of the three wave beams. The tilted  rectangle of length $L$ and width $W$ corresponds to the control area used in the model. \label{figmodele}}
\end{center}
\end{figure}

A tank $160$~cm large, $17$~cm wide is filled with $36$~cm of linearly stratified salt water with constant buoyancy frequency $N$. An internal wave of wavelength $\lambda_0$ is generated using a wave generator \citep{Gostiaux2007,Mercier2010} placed horizontally with a plane wave configuration identical to the one used by~\cite{Bourget2013}. Note that to avoid spurious emission of internal waves on the extremities of the moving region, the amplitude of the plates is constant over $n$ wavelengths in the central region, while one half-wavelength with a smooth decrease of the amplitude is added on each side. The beam width~$W=(n+1)\lambda_0$ is varied from $ \lambda_0$ to $5\ \lambda_0$ by changing the horizontal extent of the moving part of the wavemaker. A schematic view of the experimental set-up is shown in figure~\ref{figmodele}. Synthetic Schlieren technique \citep{Dalziel2000,Sutherland1999} is used to obtain the   two-dimensional instantaneous density gradient field ($\tilde{\rho}_x(x,z,t)=\partial_x(\rho(x,z,t)-\rho_0(z))$, $\tilde{\rho}_z(x,z,t)=\partial_z(\rho(x,z,t)-\rho_0(z))$)  where
$\rho(x,z,t)$ and $\rho_0(z)$ are the instantaneous and initial fluid densities. Series of experiments are performed varying the horizontal wave number ${\ell}_0$, the plate motion amplitude $a$ and the frequency $\omega_0$. It results in variations of the vertical wave number $m_0$ according to the dispersion relation $\omega_0/N=\ell_0/\kappa_0$, as well as variations of $\Psi_0$, the amplitude of the stream function $\psi$, which is defined such that $\partial_z\psi=-u$ and $\partial_x\psi=v$ with $u$ and $v$ the horizontal and vertical components of the velocity. The parameters  used in this article are summarized in Table~\ref{Table_parameters}. A typical experimental result is presented as  background of figure~\ref{figmodele}.

\begin{table}
  \begin{center}
\def~{\hphantom{0}}
  \begin{tabular}{lcccccc}
      $\textrm{Configuration}$  & \quad $N$ (rad s$^{-1})$\quad   &\quad   $\omega_0/N$\quad & \quad $\ell_0$ (m$^{-1}$) \quad &  \quad$\Psi_0/\nu$ \quad& \quad$W$ \quad&\quad\textrm{Approach}\\[3pt]
      \hline
       I    &\quad 0.89  \quad&\quad 0.85  \quad&\quad 110  \quad&\quad 16.9 \quad&\quad 1 - 5 $\lambda_0$  \quad&\quad \textrm{Experimental}\\
       II    &\quad 0.89  \quad&\quad 0.85  \quad&\quad 110    \quad&\quad 16.9  \quad&\quad 2 - 3 $\lambda_0$  \quad&\quad\textrm{Numerical} \\
       III    &\quad 0.91  \quad&\quad 0.74 \quad&\quad 75    \quad&\quad 33 \quad&\quad 3 $\lambda_0$  \quad&\quad\textrm{Experimental} \\
       IV    &\quad 0.91  \quad&\quad 0.74  \quad&\quad 75    \quad&\quad 33 \quad&\quad 3 $\lambda_0$  \quad&\quad\textrm{Numerical} \\
        V    &\quad 0.91  \quad&\quad 0.74  \quad&\quad 75   \quad&\quad 33 \quad&\quad 20 $\lambda_0$  \quad&\quad\textrm{Numerical} \\
  \end{tabular}
  \caption{Experimental and numerical parameters, with $N$ the buoyancy frequency, $\omega_0$ the frequency of the primary wave, $\ell_0$ the horizontal wave number, $\Psi_0=v_0/\ell_0$ the experimentally measured amplitude of the stream function in which $v_0$ is the vertical velocity, $\nu$ the viscosity and $W$ the width of the primary beam.}
  \label{Table_parameters}
  \end{center}
\end{table}

\smallskip
\subsection{Numerical  method}
In addition to the experiments, we performed $2$D direct numerical simulations with the finite elements commercial software Comsol Multiphysics. The simulations solve the incompressible continuity equation, the Navier-Stokes equation for a Newtonian fluid in the Boussinesq approximation and the equation of salinity conservation. 
All  elements are triangular standard Lagrange mesh of type $P_2-P_ 3$ (i.e. quadratic for the pressure field but cubic for the velocity and density fields). The total number of degree of freedom is larger than $2$ millions. At each time step, the system is solved with the Backward Difference Formulae (BDF) temporal solver and the sparse direct linear solver PARDISO. 
The Comsol BDF solver automatically adapts its scheme order between 1 and 5 (see details in~\cite{Hindmarsh2005}: it varies between 1 and 3 during our calculations. Note that no stabilization technique has been used. The Schmidt number, which compares diffusivity of salt and momentum is set to $Sc=10$ 
as a proxy for the value $Sc=700$ existing in the laboratory configuration. To prevent the creation of a reflected beam at the bottom of the domain, an attenuation  layer is added wherein viscosity and diffusivity of salt increase exponentially with depth. To mimic the wavemaker of the experimental set-up, the horizontal velocity, the vertical velocity and the density are simultaneously imposed at each time-step at the top horizontal boundary and correspond to linear gravity waves. The amplitude of the imposed velocities and density is constant over $n$ wavelengths in the central region, while one half-wavelength with a smooth decrease of the amplitude is added on each side similar to the experimental configuration. On the left and the bottom of the domain, we impose no stress and no flux and on the right, the pressure anomaly is fixed to zero, with no viscous stress and flux. Note that at time $t=0$, the linear viscous beam is imposed in the bulk. The advantage of the numerical approach is that the beam width can be varied to much larger value than~
five wavelengths.
\smallskip
\subsection{ Results.}
\begin{figure}
\begin{center}
\includegraphics{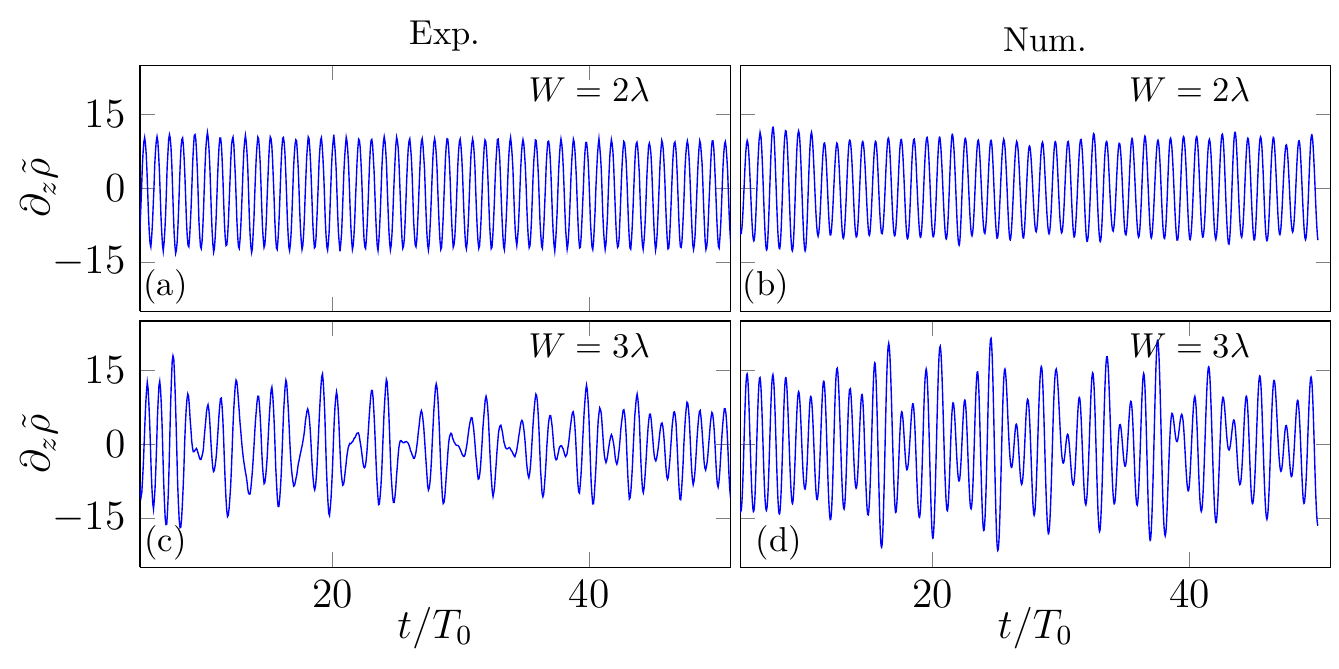}
\caption{Time evolution of the amplitude of $\partial_z\tilde\rho$ [$\textrm{kg}.\textrm{m}^{-4}$] $4.5$~cm below the wave maker and in the middle of the beam as the width of the beam is increased from $W=2\lambda_0$ (upper panels) to $3\lambda_0$ (lower panels) for configurations
 I-II ($\omega_0/N = 0.85$, $\ell_0 = 110$~m$^{-1}$, $\psi_0/\nu = 16.9$). Experimental results are shown in left panels and numerical ones in right panels.\label{figtimeseries}}
\label{drhot}
\end{center}
\end{figure}

Let us first focus on the experimental and numerical results of configurations
 I and~II. 
Figures~\ref{figtimeseries}(a) and \ref{figtimeseries}(c) present the time evolution of the vertical density gradients measured at a given point in the experiment (configuration
 I), while figures~\ref{figtimeseries}(b) and \ref{figtimeseries}(d) show the same information for the numerical simulation (configuration
 II). Experimental and numerical results display a good agreement, evidencing the following observation : while for a beam width  $W=2\lambda_0$ (figures~\ref{figtimeseries}(a,b)), the regular signal does not show any
triadic resonance, in the case where $W=3\lambda_0$ (figures~\ref{figtimeseries}(c,d)), the instability develops, as emphasized by the modulation of the signal, typical of the apparition of new frequencies in the system. Moreover, frequencies and wavelengths of the secondary waves present a good agreement, as can be seen in Table~\ref{Table_secondary_waves}.
 For experiments with $W=4\lambda_0$ and $5\lambda_0$ (not shown), the instability is observed as well.  Experimental and numerical results thus reveal for the first time the critical role of the beam width in the occurrence of the triadic instability.

\begin{table}
  \begin{center}
\def~{\hphantom{0}}
  \begin{tabular}{lcccccc}
     $\textrm{Configuration}$  &\quad  $W$ (rad s$^{-1})$\quad   &\quad    $\omega_1/N$\quad & \quad  $\kappa_1$ (m$^{-1}$) \quad &  \quad $\omega_2/N$ \quad& \quad  $\kappa_2$ (m$^{-1}$) \quad&\quad \textrm{Approach}\\[3pt]
      \hline
        I    &\quad $3-5\ \lambda_0$  \quad&\quad 0.64 \quad&\quad 220  \quad&\quad 0.25 \quad&\quad 120  \quad&\quad  \textrm{Experimental}\\
        II    &\quad $3\ \lambda_0$  \quad&\quad  0.60 \quad&\quad 201    \quad&\quad 0.26   \quad&\quad 101  \quad&\quad\textrm {Numerical}\\
        III    &\quad 3 $\lambda_0$  \quad&\quad 0.49 \quad&\quad 208   \quad&\quad 0.26 \quad&\quad 121  \quad&\quad \textrm{Experimental} \\
        IV   &\quad 3 $\lambda_0$  \quad&\quad 0.49  \quad&\quad 232    \quad&\quad 0.25 \quad&\quad 148  \quad&\quad \textrm{Numerical} \\
         V    &\quad 20 $\lambda_0$\quad&\quad 0.50  \quad&\quad 147   \quad&\quad 0.27 \quad&\quad 61  \quad&\quad \textrm{Numerical} \\
  \end{tabular}
  \caption{Experimental and numerical values of $\omega_1$ and $\omega_2$, the frequencies of the secondary waves and $\kappa_1$ and $\kappa_2$, the wave vector modulus of the secondary waves.}
  \label{Table_secondary_waves}
  \end{center}
\end{table}

\smallskip
\section{Theory} 
To understand this observation, it is crucial to realize that the theory of PSI
 is derived for infinitely extended plane waves, as described in~\cite{KoudellaStaquet2006} and~\cite{Bourget2013}. We thus propose to take into account the {\it width} of the primary wave beam, the guiding idea being that the two secondary plane waves can exit the spatial
 extent of the primary wave beam~\citep{McEwan1977,Gerkema2006}. Once they have left  this region, they cannot interact any more with the primary plane wave and the energy transfer is broken.

To perform an energy balance, let us define a control area within the primary wave beam as presented by the tilted rectangle in figure~\ref{figmodele}. For simplicity
 , we neglect the spatial attenuation of the waves in this area, i.e.  the surface energy densities $E_0$, $E_1$ and  $E_2$ of the different waves are considered uniform. Consequently, the temporal variation of the primary plane wave energy in the domain is due to: 

\noindent i) nonlinear interactions that transfer energy from the primary wave to both secondary ones (denoted $\Gamma_{\rm int}(i,j)$ where $i$ and $j$ represent the two other waves of the triad) ;

\noindent ii) viscosity; 

\noindent iii) incoming and outgoing flux of the primary wave. \\

For the domain $W\times L$, this can be summarized as

\begin{equation}
\frac{{\rm d}E_0}{{\rm d} t}LW  =\Gamma_{\rm int}(1,2)-\nu \kappa_0^2E_0LW+ E_{\rm in}W{\textbf c_{g,0}}-E_0W{\textbf c_{g,0}} \ ,
\end{equation}

\noindent with $E_{\rm in}$ the surface energy injected by the generator and $c_{g,0}=(N^2-\omega_0^2)^{1/2}/\kappa_0$.

Assuming that PSI occurs everywhere in the beam, the outgoing energy flux of the secondary waves through the cross beam faces with width W is compensated by the incoming one. In contrast, since they do not propagate parallel to the primary beam, they exit the control area also from the lateral boundaries without compensation.
For the temporal variation of the secondary waves energy in the domain, this leads to
\begin{equation}
\frac{{\rm d}E_i}{{\rm d}t}LW  =\Gamma_{\rm int}(0,j)-\nu \kappa_i^2E_iLW-E_iL |{\textbf c_{g,i}}\cdot{\textbf e_{k_0}}| \ ,
\end{equation}

\noindent with $i=1,2$, $j=2,1$ and the modulus of the group velocity $c_{g,i}=(N^2-\omega_i^2)^{1/2}/\kappa_i$.

The first term represents the interaction with the other plane waves of the triadic resonance while the third one 
accounts for the energy exiting the control area.
As $E_i\propto\kappa_i^2 \Psi_i\Psi_i^*$, using the infinite width expression~\citep{Bourget2013} for the interaction terms,
one gets
\begin{eqnarray}
\frac{{\rm d}\Psi_0}{{\rm d}t} & =&-|I_0|\Psi_1\Psi_2-\frac{\nu}{2} \kappa_0^2\Psi_0+F ,\label{equation1}\\
\frac{{\rm d}\Psi_1}{{\rm d}t} & =&+|I_1|\Psi_0\Psi_2^*-\left(\frac{\nu}{2} \kappa_1^2+\frac{|{\textbf c_{g,1}}\cdot{\textbf e_{k_0}}|}{2W}\right)\Psi_1, \label{equation2}\\
\frac{{\rm d}\Psi_2}{{\rm d}t} & =&+|I_2|\Psi_0\Psi_1^*-\left(\frac{\nu}{2} \kappa_2^2+\frac{|{\textbf c_{g,2}}\cdot{\textbf e_{k_0}}|}{2W}\right)\Psi_2 \ , \label{equation3}
\end{eqnarray}
\noindent with $I_i$ the interaction term defined as follows
\begin{equation}
I_i =\frac{\ell_j m_r - m_j \ell_r}{2\omega_i\kappa_i^2}\left[\omega_i(\kappa_j^2 - \kappa_r^2)+\ell_i N^2\left(\frac{\ell_j}{\omega_j}-\frac{\ell_r}{\omega_r}\right) \right]
\label{eq:I}
\end{equation}
with $i,j,r=0,1,2$ 
 and $F={\textbf c_{g,0}}\left(\Psi_{\rm in}^*\Psi_{\rm in}-\Psi_0^*\Psi_0\right)/({2L\Psi_0^*})$ a forcing term, corresponding to the difference between the incoming energy and the outgoing one for the primary wave.

The solution for $\Psi_1$ and $\Psi_2$ can be easily obtained with the hypothesis $\Psi_0$ constant. One finds
exponentially growing solutions with growth rate 
\begin{equation}
\sigma_{\pm}=-\frac{1}{4}\left(\Sigma_1+\Sigma_2\right)\pm\sqrt{\frac{1}{16}\left(\Sigma_1-\Sigma_2\right)^2+|I_1||I_2||\Psi_0|^2}, \label{taux_croissance}
\end{equation}
where $\Sigma_i=\nu\kappa_i^2+\sigma_{\rm adv}(\kappa_i)$ and 
$\sigma_{\rm adv}(\kappa_i)={|{\textbf c_{g,i}}\cdot{\textbf e_{k_0}}|}/{W}$, 
the inverse of an advection time since it characterizes the transport of the secondary waves energy out of the interaction region. The viscous part of the expression of $\Sigma_i$ is similar to the expression obtained in~\cite{McEwan1977} in the limit of large Prandtl number.

Cases with large $\sigma_{\rm adv}$ values will present a very strong finite-width effect.
Three parameters impact the value of the growth rate:

\begin{enumerate}
\item The width $W$ of the beam. The term $\sigma_{\rm adv}$ varies like the inverse of $W$, thus the growth rate is decreased 
when the beam is narrow. On the contrary, when $W$ goes to infinity, $\sigma_{\rm adv}$ vanishes and the growth rate of the original theory \citep{Bourget2013} is 
recovered.
\item The amplitude of the group velocities of the secondary waves. When the modulus of the secondary wave vector $\kappa_i$ is small, the group velocity increases and the secondary waves exit the primary wave beam more rapidly. So there is less time for them to grow in amplitude. Besides, $\sigma_{\rm adv}$ is weaker for low stratification~$N$ (small group velocity) 
at $\kappa$ and $\omega$ fixed, which implies a stronger instability in the oceans where $N$ is weak compared with laboratory conditions.
\item The direction of the group velocity of each waves of the triad. When the secondary waves are almost perpendicular to the primary plane wave, they leave the interaction area quickly, which does not favor the instability. To summarize, this finite-width effect is stronger when the group velocity of the secondary wave is aligned with ${\bf k_0}$.
\end{enumerate}


For configurations
 I and II presented in figure~\ref{drhot}, the above model predicts that the growth rate is an increasing function of~$W$. For an infinitely wide wave beam, the instability occurs for any value of the amplitude (see appendix). In contrast, for finite width beams, a non vanishing threshold appears. For example, for $W=2\lambda_0$, the maximum growth rate becomes negative for $\Psi_0/\nu<4.4$.  Moreover, during the finite duration of the experiment, even if the instability occurs, the amplitude of the two secondary waves might be too small to be detected. For example, we can define a detection criterion by requiring that for the secondary waves to be detected, the duration of the experiment must exceed 3 times the inverse of the growth rate. With this criterion, in the case of an infinitely wide wave beam, the instability can only be observed during the experiment if  $\Psi_0/\nu>1.6$ whereas for $W=2\lambda_0$, the threshold is five times higher, $\Psi_0/\nu>8.5$. Interestingly, this value of amplitude threshold
 has the same order of magnitude as the imposed amplitude in configurations
  I and II, notwithstanding the simplifying hypothesis made by neglecting spatial attenuation in the longitudinal (viscous decay)  and transversal (imposed wave beam shape) direction of the primary wave. Therefore, the above model gives an explanation why the instability occurs in these two configurations only for $W$ larger than $3\lambda_0$.
 
This new analysis reveals a dramatic effect on the development of the triad instability which has been totally overlooked before. When the beam gets narrower, the PSI cut-off, initially due only to viscosity, is displaced towards a larger 
forcing amplitude so that at a given amplitude, the instability can be completely suppressed by decreasing the beam width.

\section{Selection of the triad.}
\begin{figure}
\begin{center}
\includegraphics[width=0.7\textwidth]{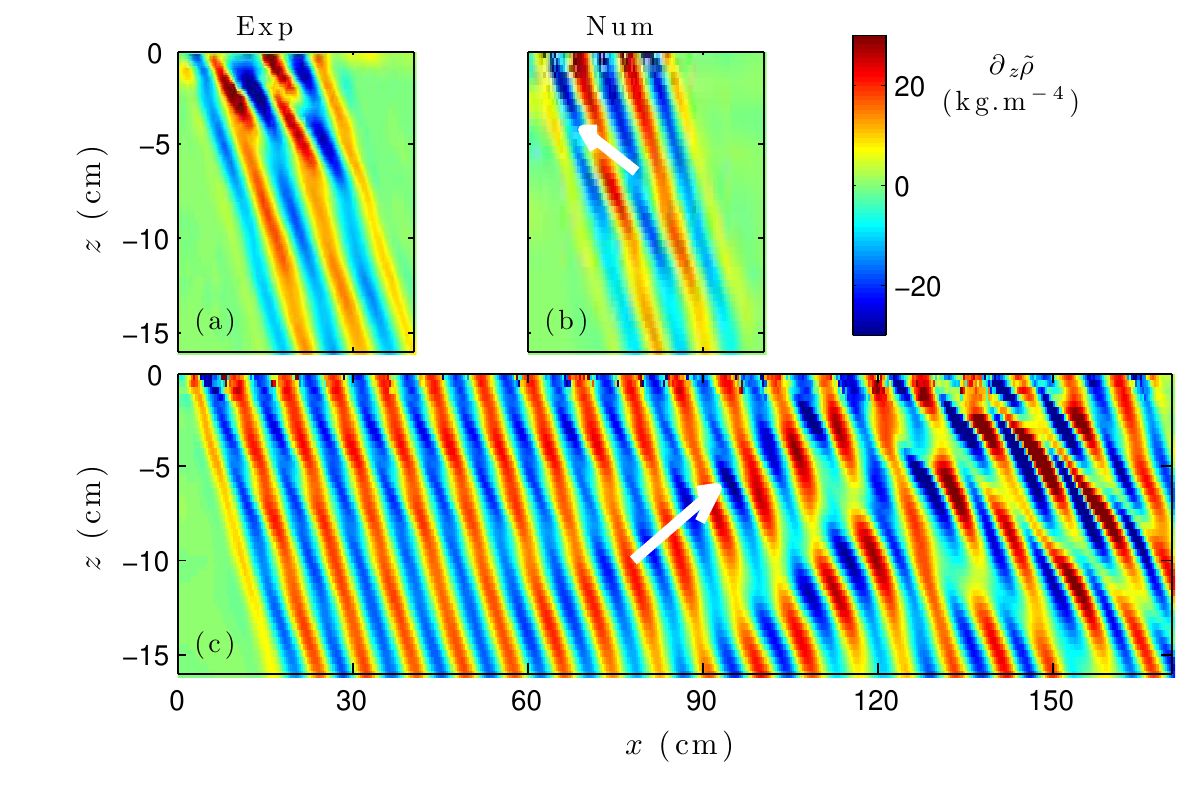}
\caption{Vertical density 
gradient fields for configurations III-V ($\omega_0/N = 0.74$, $\ell_0 = 75$~m$^{-1}$, $\psi_0/\nu = 33$). a) Experiments with $W=3\lambda_0$;  b) simulations with $W=3\lambda_0$; c) simulations with $W=20\lambda_0$. The two vectors represent the direction of the group velocity for one of the two secondary plane waves. \label{figcompa}}
\end{center}
\end{figure}

\begin{figure}
\begin{center}
\includegraphics[width=0.8\textwidth]{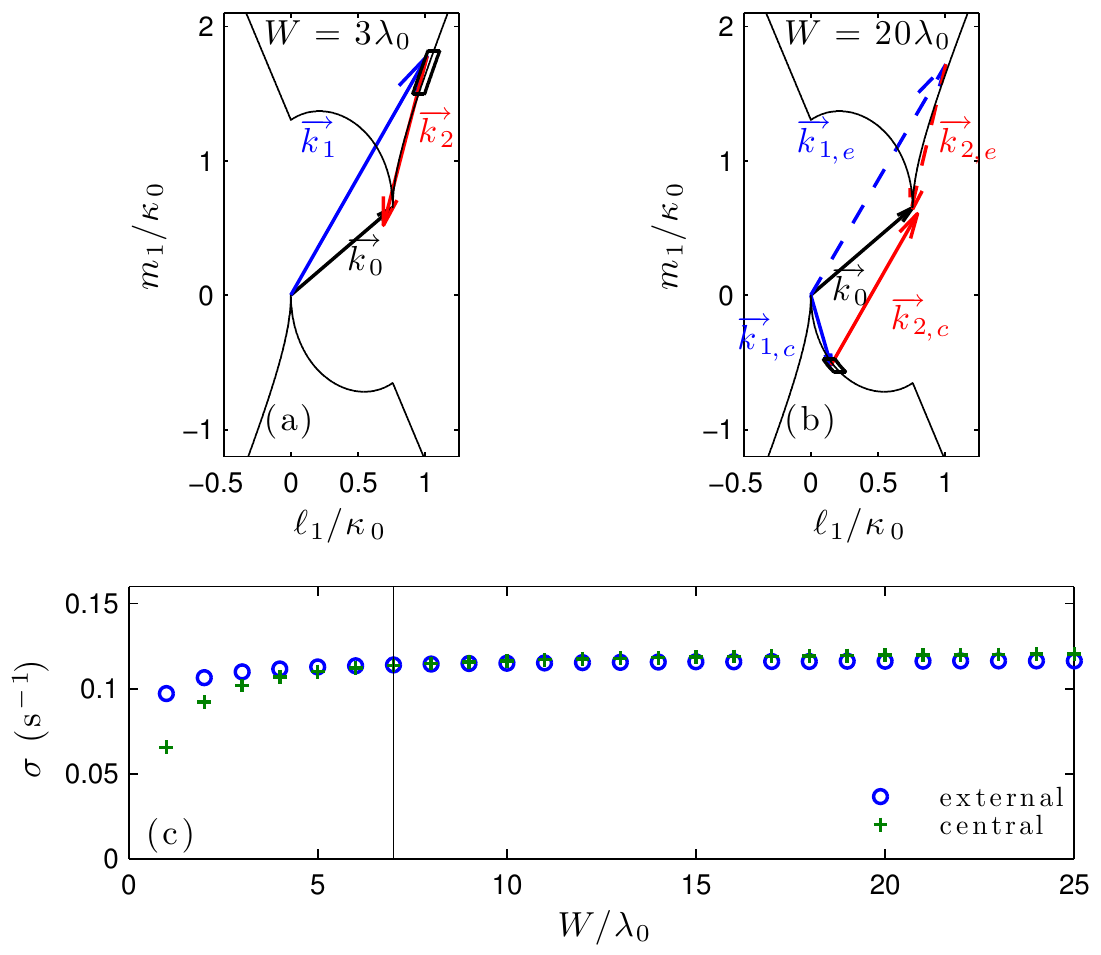}\\
\caption{(color online) a)  The three arrows are the  measurements of the three wave vectors for the numerical case $W=3\lambda_0$. b) The five arrows are the  measurements of the wave vectors for the numerical case $W=20\lambda_0$. In a) and b), the dark solid line represents the theoretical resonance loci for the secondary wave vector ${\textbf k_1}$ for a given  ${\textbf k_0}$ and the rectangles represent the most unstable mode for the {present} finite-width model. c) Evolution of the growth rate computed from (\ref{taux_croissance}) as a function 
of the width of the beam, computed using the finite-size model. The transition between the external and central triad configurations is obtained around $W=7 \lambda_0$  (vertical line).\label{figtheo}}
\end{center}
\end{figure}

We will now show that the finite-width effect can also result in a specific triad selection. To do that, we
focus on the results for another set of parameters (Configurations
 III-V, see Table~\ref{Table_parameters}), this case is therefore different from the previous one.
Experimental and numerical wave fields (respectively configurations
 III and IV) presented in figures~\ref{figcompa}(a) and \ref{figcompa}(b)  are in good agreement. As underlined by~\cite{Bourget2013}, the theoretical prediction for an infinite primary plane wave is that for this set of parameters (configurations
  III-IV), the wavelength of one of the secondary waves generated by the instability is
 larger than the primary wavelength, while the other one is smaller (see Table~\ref{Table_secondary_waves}). In this case, the energetic transfer will occur
towards larger and smaller scales simultaneously. However, this prediction was not verified and a different type of triad was observed experimentally in~\cite{Bourget2013}, with transfer to smaller scales only. 
This difference between the prediction and the observations can again be traced back to the finite width of the 
wave beam. To demonstrate that, we rely on numerical results since the experimental set-up cannot generate plane waves 
with a beam 
size larger than $6\ \lambda_0$. Note that we have performed numerical simulations for $6\lambda_0$ (yielding similar results as for $3\lambda_0$), and $10\lambda_0$ (yielding similar results as for $20\lambda_0$). Therefore, we do not show these results here and focus only on the extreme cases $3\lambda_0$ and $20\lambda_0$. Figure~\ref{figcompa}(c) shows the results of configuration
 V,   i.e. for the same parameters but for a 
significantly wider beam ($20\lambda_0$). The primary wave beam is still unstable, but the secondary waves 
look quite different compared to the results for $W=3\lambda_0$. To 
quantify the differences, a temporal Hilbert transform~\citep{Mercier:PoF:08} and a spatial Fourier transform are used to measure the different wave vectors present in the numerical density gradient fields. These vectors are shown in figures~\ref{figtheo}(a) and \ref{figtheo}(b). In addition, the curves in these figures
represent the location of the tip of all possible wave vectors ${\textbf k_1}$ satisfying the theoretical resonance conditions with a growth rate possessing a positive real part. A clear difference between the two cases is visible. For $W=3\lambda_0$, a single triad is observed and 
its secondary wave vectors ${\textbf k_1}$ belongs to an external branch of the theoretical resonance loci ({$\ell_1>\ell_0$}). The 
wavelengths of both secondary waves are smaller than the primary wavelength. For  $W=20\lambda_0$, two different triads are measured. The first one (dashed vectors), with secondary vector
${\bf k}_{1,e}$ lying on an external branch (hence the subscript ``$e$''), is similar to the one found for smaller $W$. For the second one (solid vectors), its secondary vector ${\bf k}_{1,c}$  is located in the central region of the theoretical loci 
curve ({$0<\ell_1<\ell_0$}).  In this case, one secondary 
wave has a larger wavelength than the primary wave and the other a smaller one.

These observations can be explained by the finite-width theoretical approach presented previously: the predicted evolution of the maximum value of 
the growth rate as a function of $W$ is shown in figure~\ref{figtheo}(c). The growth rates were computed separately for the external ($\circ$ symbols) and central ($+$ symbols) cases. A transition between the two possible triads is predicted around $W_c= 
7\lambda_0$. For $W<W_c$, the most unstable triad is on the external branch. The corresponding predicted location of the tip of wave vector~${\textbf k_1}$ is displayed as a rectangle in figure ~\ref{figtheo}(a), showing a very good agreement with the 
numerical observation. For this narrow beam, this agreement extends as well to the experimental data. On the contrary, for wide enough beams, i.e. above the threshold~$W_c$, in spite of the fact that ${\bf k}_{1,c}$ is almost perpendicular to ${\bf k}_{0}$ (a condition enhancing the value of $\sigma_{\rm adv}$), the secondary waves have time to grow before leaving the interaction area. Consequently, the central triad becomes dominant when 
increasing the beam width. In the time evolution of the numerical experiment, this triad appears first, the result derived for infinitely wide beams is thus recovered. At a later time, it is followed
 by the second triad. 
 Our numerical 
results confirm that the width of the beam changes the selection of the triad. This new effect
explains why energy transfer is mainly towards smaller scales for narrow beams.

\section{Oceanic case}
This new finite-width theory is therefore of importance when considering
internal wave beams from in-situ observations of the ocean~\citep{Lien2001,Dewan1998,Gostiaux2007b}. In the ocean, since the wavelengths are larger by several orders of magnitude, the Reynolds number is much larger than in the experiments and consequently viscous effects can be safely neglected. Consequently, the beam width is the dominating control parameter for the growth rate. We will focus on equatorial regions, for which the background rotation has no effect.

\begin{figure}
\begin{center}
\includegraphics[width=\linewidth]{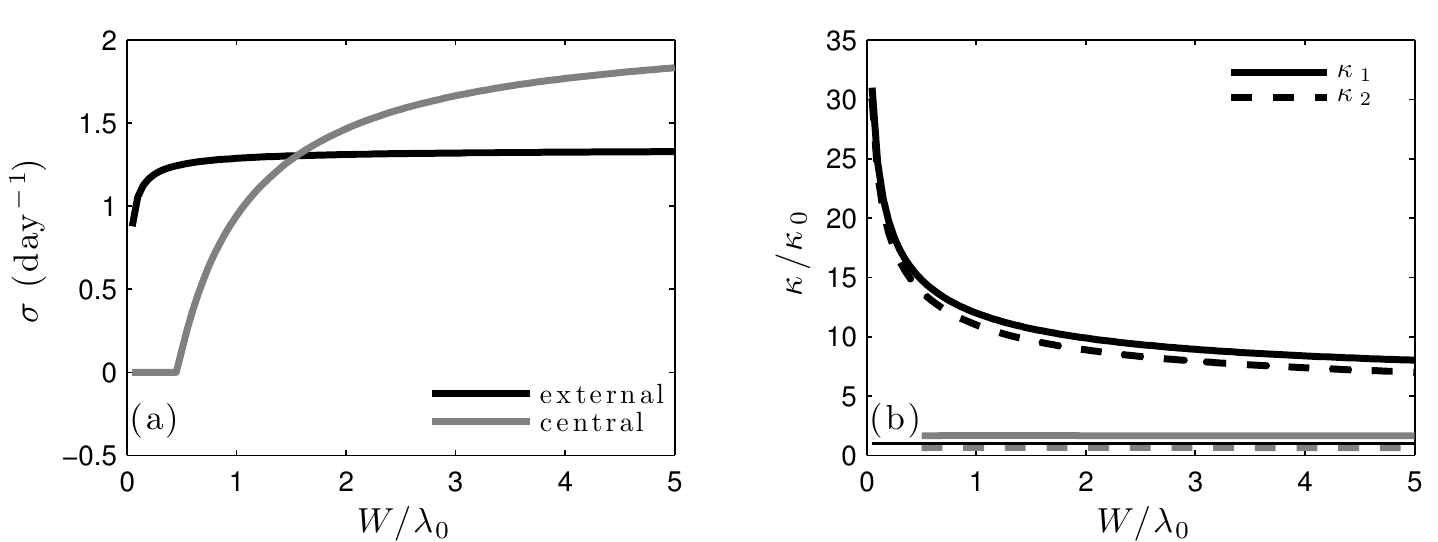}
\caption{(a) Evolution of the growth rate computed from (\ref{taux_croissance}) as a function 
of the width of the beam, computed using the finite-size model in an oceanic case ($N = 1.12\,10^{-3}$~rad/s, $\omega_0/N = 0.125$, $\lambda_0 =100$~m, $f = 0$). The transition between the external and central triad configurations is obtained around $W=1.5 \lambda_0$. (b) Evolution of the modulus of the secondary wave vectors for the external case in black and the central case in gray as a function of the width of the beam. The horizontal thin black line represents $\kappa/\kappa_0=1$.}
\label{casoceanique}
\end{center}
\end{figure}

For example, we consider a primary beam with the following typical parameters:
\begin{enumerate}
\item $\omega_0=1.4\cdot10^{-4}$ rad$\cdot$s$^{-1}$, which corresponds to the diurnal M2 tide period.
\item $\lambda_0=100$ m, which corresponds to field measurements in Kaena Ridge, Hawaii \citep{Sun2013}.
\item $N=1.12\cdot10^{-3}$ rad$\cdot$s$^{-1}$. This value of the buoyancy frequency allows a propagation angle of $\theta=7$\textdegree~which is close to oceanic observations.
\item the Froude number $Fr=u_0m_0/(2\pi N)=0.035$ \citep{Gayen2013} which allows us to estimate $\Psi_0$.
\item the vertical component of the Coriolis force is ignored and the Coriolis parameter $f$ is set to zero.
\end{enumerate}
Figure \ref{casoceanique}(a) shows the evolution of the growth rate as a function of the width of the beam. With the chosen  parameters, the transition between the external and central triad configurations is obtained around $W=1.5\lambda_0$. Figure \ref{casoceanique}(b) presents the evolution of the modulus of the secondary wave vectors for the external case (black lines) and the central case (gray lines) as a function of the width of the primary beam. For example for a narrow beam $W=1\lambda_0$, the maximum of growth rate is obtained on the external branch (\ref{casoceanique}(a)) and PSI enables a transfer to smaller scales: $\kappa_1=12\kappa_0,\ \kappa_2=11\kappa_0$ (figure \ref{casoceanique}(b)). This behavior corresponds to simulations~\citep{Gerkema2006,Gayen2013} and to oceanic observations~\citep{Mackinnon2012}. Moreover the value of the growth rate predicted by the model  ($\sigma=1.2$ day$^{-1}$) gives a value which has the right order of magnitude when compared to numerical values: 0.5 day$^{-1}$ \citep{Gerkema2006}, or 0.66 day$^{-1}$ \citep{Gayen2013} and to oceanic measurements: from 0.2 to 0.5 day$^{-1}$ \citep{Mackinnon2012}. On the contrary for a larger beam, the instability enables transfer to both larger and smaller scales ($\kappa_1=1.7\kappa_0,\ \kappa_2=0.7\kappa_0$) which corresponds to the infinitely wide theory prediction.

This transition depends on the amplitude of the stream function $\Psi_0$. For example, for $\Psi_0$ three times larger, the transition is obtained for $W=0.55\lambda_0$. In contrast, for $\Psi_0$ three times smaller, the transition is obtained for $W=4.5\lambda_0$. Thus, the transition between the two behaviors appears for $W$ close to the typical width of an oceanic beam. Consequently, the finite width of the beam can have a notable impact on the selection of the triad in oceanic cases and provides an explanation for the predominance of energy transfer to smaller scales for oceanic narrow beams.

\section{Conclusion.}
We have shown theoretically, numerically and experimentally that the width of the internal wave beam
is a key element in parametric subharmonic instability. This feature had been totally overlooked previously, despite
its dramatic consequences on the triad selection mechanism. The subharmonic plane waves that are theoretically unstable can only extract energy from the primary wave if they do not leave the primary beam too quickly. 
This finite-width mechanism has two opposite consequences on the wave energy 
dissipation: it introduces a PSI threshold (reducing transfer and therefore dissipation), but
when PSI is present it enhances the transfer towards small wavelengths, more affected by dissipation.
A complete theoretical study of the impact of the envelope on the PSI will be a timely achievement. We are aware of a recent work about the weakly nonlinear asymptotic analysis of the problem by~\cite{KarimiAkylas2014}.

It has not escaped our notice that the Coriolis force will significantly modify the prediction. 
Indeed, the group velocity for inertia-gravity waves, which is proportional to $\sqrt{(\omega^2-f^2)\cdot(N^2-\omega^2)}/(\omega\kappa)$, decreases with the Coriolis parameter $f$~\citep{Gill1982}. The rotation reducing the ability of subharmonic waves to escape, it seriously reinforces the instability. At the critical latitude, the group velocity
vanishing, one should even recover the theoretical prediction for plane waves.



Finally, from a more fundamental point of view, such a mechanism modifies significantly the transfer of energy
between scales and must be taken into account in all analysis~\citep{caillol2000,Lvov2010,Lvov2012} of wave turbulence, in which infinitely wide plane waves
are until now the common theoretical objects, but not appropriate for careful predictions.



\begin{acknowledgments}
We thank P. Meunier for insightful discussions. This work has been partially supported by the ONLITUR grant (ANR-2011-BS04-006-01)
and achieved thanks to the resources of PSMN from ENS de Lyon. MLB acknowledges financial support from the European Commission, Research Executive Agency, Marie Curie Actions (project FP7-PEOPLE-2011-IOF-298238).

\end{acknowledgments}

\section*{Appendix. Is there an amplitude threshold in resonant triadic instability for an infinitely wide wave beam?}
The expression~(\ref{taux_croissance}) shows that, to get a strictly positive growth rate, the amplitude of the stream function has to be larger than 
\begin{equation}
|\Psi_s(l_1,m_1)|=\frac{\nu}{2}\sqrt{\frac{\kappa_1^2\kappa_2^2}{I_1I_2}}\ ,
\label{Psi_S_annex}
\end{equation}
with $I_1$ and $I_2$ defined in Eq.~(\ref{eq:I}). This expression has already been reported in~\cite{KoudellaStaquet2006} and~\cite{Bourget2013} with minor typos in the latter case. What has been overlooked is that the PSI threshold is the global minimum of this function~(\ref{Psi_S_annex}) of several variables.

In this appendix, we study the behaviour of $\Psi_s$ when $\overrightarrow{k_1}$ tends to $\overrightarrow{k_0}$, which is the most unstable case for small values of the amplitude.
In this case, we assume
\begin{eqnarray}
\ell_1&=&\ell_0(1+\mu_0\epsilon^{\alpha})\ , \\
m_1&=&m_0(1+\epsilon)\ ,
\end{eqnarray}
with $\epsilon=o(1)$, $\alpha\geqslant1$ and  $\epsilon$ and $\mu_0$ are positive.

Using the temporal and spatial resonance conditions and the dispersion relation of internal waves, we obtain the relation
\begin{equation}
\frac{m_0^6}{\kappa_0^6}\left(\epsilon^4-2\mu_0\epsilon^{3+\alpha}+o(\epsilon^{2\alpha})\right)=\mu_0^2\epsilon^{2\alpha}\ .
\end{equation}
To solve this equation, there are {\it a priori} two different solutions for $\alpha$:
\begin{enumerate}
\item $3+\alpha=2\alpha$ leading to $\alpha=3$. However, this value makes it possible to balance terms at order $\epsilon^6$ but not the lower order term $\epsilon^4$. This value $\alpha=3$ is therefore not acceptable.
\item $4=2\alpha$ leading to $\alpha=2$. In that case, the lowest order terms can be balanced and one gets $\mu_0=(m_0/\kappa_0)^3$.
\end{enumerate}
Finally, we obtain
\begin{eqnarray}
\ell_1&=\ell_0(1+\mu_0\epsilon^{2})  \quad \textrm{and} \qquad \ell_2&=-\mu_0\ell_0\epsilon^2 \\
m_1&=m_0(1+\epsilon) \  \qquad \textrm{and} \quad\  \ m_2&=-m_0\epsilon\,.
\end{eqnarray}
With these relations, it can be shown that 
\begin{equation}
I_1=-\ell_0m_0\epsilon+o(\epsilon) \qquad \textrm{and} \qquad I_2=-\ell_0m_0+o(1)\ ,
\end{equation}
which means that
\begin{equation}
|\Psi_s|=\frac{\nu}{2}\frac{N}{\omega_0}\sqrt{\epsilon}+o(\epsilon^{1/2})\ .
\end{equation}
Therefore, the minimum of the positive expression~(\ref{Psi_S_annex}) is zero. Consequently, there is no threshold for an infinitely wide wave beam.

\end{document}